# A simple, picojoule-sensitive ultraviolet autocorrelator based on two-photon conductivity in sapphire

KENNETH J. LEEDLE,[1,*] KAREL E. URBANEK,[2] ROBERT L. BYER[2]

[1]Department of Electrical Engineering, Stanford University, Stanford, CA 94305, USA.
[2]Department of Applied Physics, Stanford University, Stanford, CA 94305, USA.
*Corresponding author: kleedle@stanford.edu

December 19, 2016

**We present a simple autocorrelator for ultraviolet pulses based on two-photon conductivity in a bench-top fabricatable sapphire sensor. We perform measurements on femtosecond 226 - 278 nm ultraviolet pulses from the third and fourth harmonics of a standard 76 MHz titanium sapphire oscillator and picosecond 266 nm pulses from the fourth harmonic of a 1064 nm 50 MHz neodymium vanadate oscillator. Our device is sensitive to 2.6 pJ ultraviolet pulses with peak powers below 20 W. These results represent the lowest measured autocorrelation peak powers by over one order of magnitude for a system with no reference pulse in the deep ultraviolet ( < 300 nm). The autocorrelator can potentially support UV pulse lengths from 50 fs – 10's of picoseconds.**

## 1. INTRODUCTION

Ultrafast laser pulses in the deep ultraviolet (UV) are important for a number of applications, from spectroscopy to photoemission studies [1]. Ultraviolet pulses are readily available in many laboratories, but the characterization of these pulses is much more involved than their production even after several decades. The second harmonic generation crystals that are routinely available for visible or infrared autocorrelation or frequency resolved optical gating are not available for ultraviolet pulses, complicating their characterization. In particular, the measurement of the ultraviolet harmonics of standard laser oscillators has remained challenging due to the limited peak power available. The ability to characterize non-amplified deep ultraviolet pulses with a simple setup is critical for high repetition-rate experiments in applications like electron photo-emission and spectroscopy.

There have been a variety of techniques demonstrated to measure ultraviolet pulse lengths: sum-frequency or difference-frequency generation with a reference pulse, two-photon absorption (TPA) in bulk materials, or two-photon absorption in specialized detectors. Sum-frequency or difference-frequency generation with a well characterized reference pulse can characterize pulses with watt-level peak powers, but this requires synchronization and characterization of the reference pulse, and is subject to phase-matching limitations [2]. Additionally, frequency mixing schemes are cumbersome for practical systems since the temporal overlap has to be reestablished after any modification.

Two-photon absorption autocorrelation in bulk materials like diamond [3] and BBO [4] is best suited to amplified or high-power laser systems with multi-nJ UV pulse energy or over ~50 kW peak UV power. Other methods like self-diffraction and transient grating frequency resolved optical gating are only applicable for very energetic pulses [5].

Specialized UV two-photon detectors have been demonstrated based on TPA in a diamond pin photodiode [6] or a CsI photomultiplier tube (PMT) [7]. These devices had peak power sensitivities of 19 kW and 500 W, respectively, for 265 nm and 300 nm pulses. At lower pulse powers these devices are limited by parasitic linear absorption (which limited the PMT technique to wavelengths > 300 nm). The fourth and fifth harmonics of a cavity-dumped Nd:YAG mode-locked laser were measured down to 950 W peak power using a custom-built PMT with low parasitic absorption [8]. Additionally, a planar fused silica photoconductive switch based on interdigitated electrodes has been demonstrated with 10 nJ energy or 76 kW peak power sensitivity at 267 nm [9]. Due to the device design, however, this two-photon conductive sensor experienced high dark current, which limited its sensitivity. The fused silica detector had widely spaced electrodes as well, which could not take advantage of the micron-size spots that are readily achievable in the UV band for maximum sensitivity. Additionally, the diamond pin photodetector and fused silica photoconductive switch require complex fabrication facilities, limiting the widespread use of those devices as they are not commercially available yet.

In this letter, we present a UV autocorrelator based on two-photon induced conductivity in sapphire that has picojoule, sub-20 W peak power sensitivity at 267 nm, has sub-1pA dark current, essentially zero parasitic linear absorption, and is easily fabricated. The nonlinear device is based on a silicon-sapphire-silicon stack, uses only off-the-shelf parts, and can be benchtop-fabricated in a few minutes with no specialized equipment. The high sensitivity of the sapphire two-photon conductivity (TPC) sensors allows them to work with UV pulse energies readily accessible with the harmonics of low power laser oscillator systems. This is, to the best of our knowledge, the first published pulse length measurement of a sub-300 nm laser pulse in a non-amplified CW mode-locked laser with no reference pulse.

## 2. DEVICE DESIGN AND CHARACTERIZATION

The sapphire two photon conductive sensor design is shown in Figure 1. It uses a piece of 1-4 ohm-cm phosphorus-doped silicon as the base electrode. A 25±10 μm thick A-plane sapphire piece (Valley Design Corp.) is set on top of the base electrode, and a cleaved piece of 1-4 ohm-cm phosphorus doped silicon is used as the top electrode. The cleaved edge of the top silicon electrode forms a very sharp taper out to the edge. The assembly is held together with polyamide tape. This device design was chosen for its simplicity and ease of fabrication. Silicon was chosen for the top and bottom electrodes because it is electrically conductive, readily available in optically polished wafers, can be easily cleaved to form very sharp edges, and is very stable even under high fields. The base electrode is biased to -480 V, for an approximately 19.2 kV/mm electric field across the gap. The devices were stable under these high electric fields with no degradation of performance over several months of use. The -480 V bias is comparable to bias voltages routinely used with photomultiplier tubes.

The UV pulses are focused onto the edge of the top electrode and a photo-induced conductive channel is formed through the sapphire to the base electrode. The current collected on the top electrode is measured with a calibrated transimpedance amplifier. Optical absorption in the bulk silicon electrodes does not result in continuous current flow; the device's dark current without UV illumination of the electrode gap is about 0.7 pA. Alternatively, metalized sapphire electrodes could be used but these introduced significant dark current in our tests.

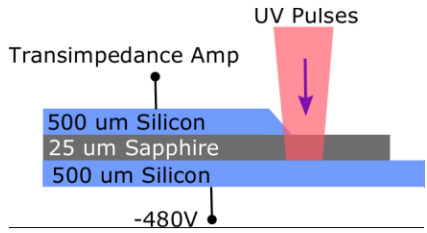

Figure 1. Two-photon conductive sapphire sensor illustration.

Femtosecond UV pulses were generated from a Coherent MIRA 900 tunable titanium sapphire oscillator using sum-frequency generation of the fundamental output and its second harmonic. This third harmonic output was tunable from roughly 250 nm to 300 nm, limited by phase-matching in the nonlinear crystals and group delay compensation. The repetition rate of the MIRA was 76 MHz. The Ti:Sapphire fundamental was doubled in 1 mm thick θ = 29.2° type-I beta-barium borate (BBO) crystal, delay compensated with a calcite plate, the fundamental and second harmonic were aligned with a dual waveplate, and their sum-frequency was generated in 1.5 mm θ = 47° type-I BBO. The total UV generated was 0.1 to 2.2 mW depending on wavelength. The sapphire TPC sensor's short wavelength limits were tested with the fourth harmonic of the MIRA. The second harmonic MIRA pulses were again frequency doubled in 200 μm thick θ = 44.3° type-I BBO at a steep angle of incidence. This allowed generation of 210 - 226 nm UV pulses at 40 – 300 μW average power. Additionally, picosecond 266 nm pulses were generated from the fourth harmonic of a High Q Picotrain neodymium vanadate 50 MHz oscillator. This system was doubled in a 9 mm type-II potassium titanyl phosphate (KTP) crystal and then doubled again in 3 mm thick θ = 47° type-I BBO to produce 10 mW of 266 nm light.

The UV beams were focused onto the device using an f = 25 mm off-axis parabolic mirror (OAP) to a spot size of approximately 3.4 x 11 ±0.4 microns 1/e$^2$ radius as measured with a knife edge. The UV spot was not diffraction limited due to scattering on the diamond-turned mirror surface. The focused beam's Rayleigh length of 140 μm was sufficient to propagate through the 25 μm sapphire plate. The focused beam was positioned adjacent to the edge of the top sapphire plate to achieve maximum two-photon induced current in the sapphire TPC sensor.

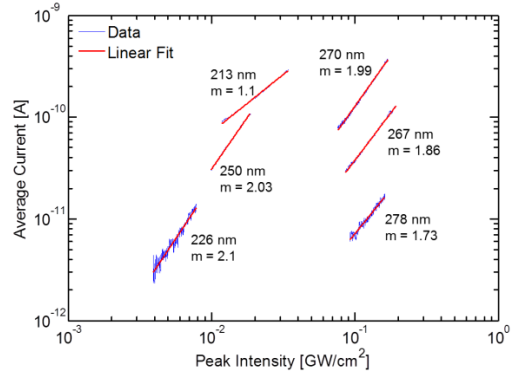

Figure 2. Log-log plot of the average photoconductive current of the sapphire TPC sensor as a function of incident laser intensity for several UV wavelengths. Data and a log-log linear least-squares fit (m = slope) are shown. A device bias voltage of -480V was used.

Figure 2 shows the average photoconductive current of the silicon-sapphire sandwich sensor as a function of the incident laser intensity on a log-log scale. As can be seen, a slope of roughly 2 is shown for wavelengths from 226 nm to 278 nm. This confirms that the sapphire sensor photoconductivity is quadratic with the incident laser intensity, as expected for a two-photon absorption device. The responsivity was not measurably affected by the UV laser polarization. Two photon conductivity measurements below 225 nm showed linear intensity dependence, as shown in Figure 2 for the case of 213 nm. This likely stems from defects or impurities in the standard-grade sapphire used [10]. Vacuum UV grade sapphire could possibly be used in devices for characterizing pulses in the range from 150 - 225 nm. Despite there being linear absorption in the sub-225 nm range for our device, there was no measurable two-photon photocurrent measured using the second harmonic of the Ti:sapphire laser in the 375 - 450 nm range even at ~10 GW/cm$^2$ intensities. This is presumably due to the relatively low number of defects in the sapphire lattice.

Following [9], we calculate the two-photon absorption coefficient β of sapphire from,

$$\beta = \frac{4N_q W_{ph} \tau_p}{ALF^2} \quad \textbf{(2.1)}$$

where $N_q$ is the number of electrons produced via TPA, $W_{ph}$ is the UV photon energy, $\tau_p$ is the laser pulse duration, A is the laser spot cross-sectional area, L is the device layer thickness, and F is the laser fluence.

Figure 3 shows the two-photon responsivity of the sapphire TPC sensor as a function of wavelength. The TPA coefficient varies from roughly 0.15 cm/GW at 226 and 250 nm to approximately 0.01 cm/GW around 266 nm, and drops off beyond 278 nm. This energy cutoff agrees well with half the 8.9 eV bandgap energy of sapphire [11]. The uncertainty in the TPA coefficients is due to the laser spot size uncertainty. The TPA coefficients at 266 nm are roughly one order of magnitude below the sapphire TPA coefficients measured via z-scan measurements [12]. This reduction in the measured current extraction is likely due to carrier recombination, and is similar to the effect seen in reference [9]. Cryogenic temperatures have been shown improve charge transport in sapphire by orders of magnitude [13], and could be employed for higher sensitivity.

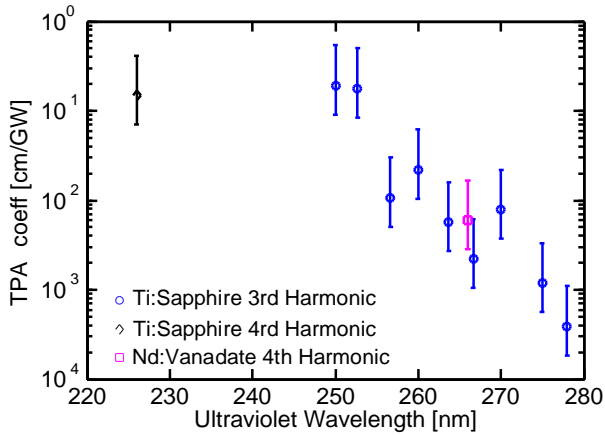

Figure 3. Two-photon absorption coefficient β as a function of ultraviolet wavelength for the silicon-sapphire sandwich device.

## 3. AUTOCORRELATION MEASURMENTS

Noncollinear autocorrelation measurements were performed using a Michelson interferometer design as shown in Figure 4. Bare aluminum hollow retroreflectors with approximately 50% UV throughput were used, and the beamsplitter had a 1 mm thick UV fused silica substrate. The beamsplitter did introduce some dispersion, but this was not significant for our 150 fs UV pulses. The total optical throughput of the autocorrelator was about 20% from the input to the sapphire sensor due to mirror and beamsplitter losses. Split mirror and other autocorrelator designs can have much lower losses and less dispersion [14], but this design was chosen for its simplicity and ability to accommodate pulse widths from ~50 femtoseconds to 10's of picoseconds. The sapphire TPC sensor only adds approximately 7.6 fs$^2$ of dispersion at 267nm, making it compatible with <10 fs pulses with an all-reflective autocorrelator design.

Noncollinear autocorrelation with crossed beams in the two-photon sensor was found to have the best signal to noise ratio for a 0.5 Hz acquisition rate with our setup. This was in part due to the low bandwidth of the transimpedance amplifier and the fact that our pulses were chirped. Autocorrelation peak-to-background ratios of 1.8 to 2.1 were typical for the autocorrelator. This is presumably due to imperfect overlap of the focused beams inside the sapphire. This is consistent with the peak-to-background ratios obtained with other photoconductive or photoelectric sensors in the UV [7-9].

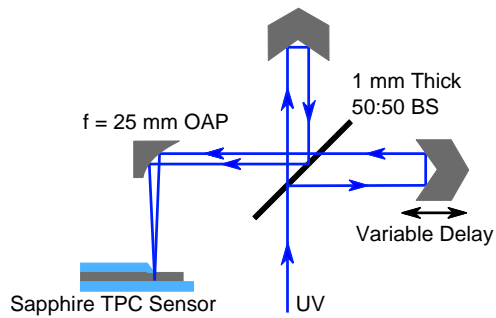

Figure 4. Schematic of the Michaelson non-collinear UV autocorrelator with sapphire TPC sensor.

Figure 5(a) shows a typical autocorrelation trace of the third harmonic of the titanium sapphire laser using the sapphire TPC sensor and 2.0 mW of 266.7 nm light (180 W peak power). The measured full-width at half maximum of the autocorrelation trace was 224 fs, corresponding to a sech$^2$ pulse width of 146 fs. These pulses were chirped coming out of the third-harmonic generator due to several lenses used in the beam path. Pulse stretching measurements by adding known amounts of dispersion indicate that the third harmonic UV pulses could be compressed to approximately 50 fs. An UV spectrometer was not available to calculate the time-bandwidth product, and pulse compression was not attempted. Figure 5(b) shows a typical autocorrelation trace of the 266 nm fourth harmonic of the Nd:vanadate laser using 7.0 mW average power (28 W peak power). The measured full-width at half maximum of the autocorrelation trace was 7.6 ps, corresponding to a sech$^2$ pulse width of 5.0 ps. The time-bandwidth product of these picosecond pulses was also not calculated due to the lack of a suitable UV spectrometer.

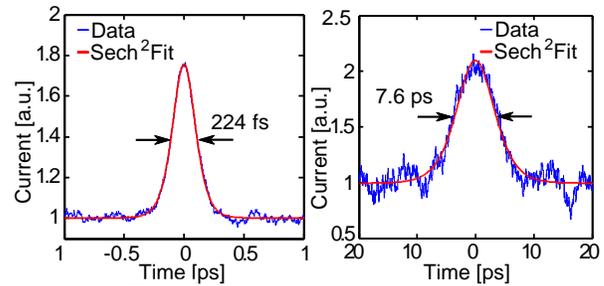

Figure 5. (a) Typical non-collinear autocorrelation trace of the third harmonic of the Ti:sapphire laser and sech$^2$ fit using the sapphire TPC sensor. 2 mW of 267 nm were used, and the measured sech$^2$ pulse length is 146 fs. (b) Non-collinear autocorrelation trace of the fourth harmonic of the picosecond Nd:Vanadate laser with 7 mW UV power, with a 5.0 ps sech$^2$ pulse width.

The sapphire TPC autocorrelator was able to measure 267 nm, 146 fs pulses with average powers down to 200 μW at 76 MHz, or 2.6 pJ per laser pulse with a peak power of 18 W. Using the fourth harmonic of the picosecond Nd:vanadate laser, the autocorrelator could measure pulses down to 4 mW average power for the 50 MHz, 5.0 ps pulses, corresponding to a peak power of 16 W. This approaches the 2.5 W sensitivity achieved with cross-correlation measurements [2], with a small fraction of the complexity.

## 4. CONCLUSION

In summary, we have demonstrated an easily fabricatable two-photon conductive sensor based on sapphire for ultraviolet autocorrelation measurements. The two-photon sensor had a quadratic response to UV light from 226 – 278 nm, and the autocorrelator could measure picojoule femtosecond to picosecond pulses with peak powers below 20 W. This enabled the first direct pulse-length measurements of the UV harmonics of standard mode-locked laser oscillators without amplification or frequency mixing. Future improvements could include using vacuum UV grade sapphire for shorter wavelength compatibility, cryogenic cooling of the sapphire sensor for better charge transport, improved focusing mirrors for a tighter focus onto the sensor, and improved hollow retroreflectors and beamsplitter for higher reflectivity and lower dispersion. Additionally, a different two photon material with a smaller bandgap could be used to measure wavelengths longer than 278 nm.

**Funding.**

Moore Foundation No. 4744; Air Force Office of Scientific Research FA9550-14-1-0190.

**Acknowledgement.**

The authors are grateful to Carsten Langrock, Marc Jankowski, Martin M. Fejer, and Andrew Ceballos for helpful discussions.